# Computational KJ-Hō: An Analyst-Bias-Free Insight Extraction Framework from Large-Scale Qualitative Data Using Domain-Specialized LLMs

**Kasumi Ban**

Sophia University Graduate School, Graduate Program in Applied Data Sciences
kajungbang@sophia.ac.jp

## Abstract

The qualitative research methodologies that have long underpinned the generation of consumer insights—the KJ method, Grounded Theory, and Thematic Analysis—are all subject to a fundamental constraint: the cognitive processing capacity of the human analyst. Even a skilled researcher can handle only tens to a few hundred data units within a practical timeframe. In mature consumer markets where individual needs grow increasingly heterogeneous, this ceiling prevents practitioners from grasping the holistic structure of consumer psychology at the scale that contemporary data collection makes possible, turning it into a strategic bottleneck. Moreover, replication studies in the social and behavioral sciences have shown that conclusions can vary substantially across analysts even for identical data (analyst bias); simply increasing the number of analysts therefore tends to amplify this bias rather than mitigate it.

Over the past two years, numerous end-to-end automation systems have been proposed, including LOGOS, Neo-Grounded Theory, and AcademiaOS. These systems, however, share five common limitations: (1) reliance on general-purpose LLMs that lack the internalization of domain knowledge through Continued Pre-Training (CPT); (2) analysis confined to a single corpus, failing to capture cross-interview segment patterns; (3) the absence of evaluation metrics tailored to the domain; (4) a disconnect between analytical output and actionable strategy; and (5) an overreliance on Western qualitative research traditions, accompanied by the absence of non-Western methodologies such as the KJ method.

This study proposes Computational KJ-Hō (the Kawakita Jiro method), a theoretical framework that computationally realizes the epistemology of the KJ method—allowing structure to emerge from the data itself without imposing the analyst's preconceptions. We term this orientation "analyst-bias-free." The framework employs a domain-specialized LLM that undergoes CPT on a marketing-research corpus, followed by Supervised Fine-Tuning (SFT) on expert-curated insight pairs, and is organized as a three-layer architecture comprising data structuring, insight extraction, and strategy generation. Two preliminary studies conducted in the Japanese marketing context show that this conceptual pipeline produces analytically meaningful output, while also showing that general-purpose models, as they stand, fall short of the semantic distinctions drawn by skilled researchers—supporting the necessity of CPT-based domain specialization. The framework does not remove the human analyst: the analyst retains a supervisory role, setting the analytical objectives and evaluating outputs for strategic decision-making. This paper is a concept paper that submits the theoretical framework to scrutiny ahead of empirical validation.

This paper makes five contributions: (1) a novel theoretical framework that integrates the KJ method, Grounded Theory, and Peircean abductive reasoning into a single epistemological commitment of "data-driven explanation generation”; (2) a three-layer architecture design that leverages domain-specialized embeddings for cross-interview analysis; (3) two novel metrics—InsightExtraction-F1 and MarketingQA—for evaluating insight quality in domain-specific contexts; (4) an explicit engagement with the WEIRD problem in AI-driven qualitative research, placing a non-Western methodology at its core; and (5) five practice-derived problem formulations drawn from nearly three decades of global marketing-research experience, translated into design requirements that integrate domain expertise with computational methodology.

# 1 Introduction

## 1.1 The Scalability Crisis in Qualitative Research

Consider a typical scenario in marketing-research practice. A team is commissioned to portray the full picture of consumers' behavior and attitudes toward a newly launched ready-to-drink (RTD) beverage. They recruit 40 interviewees—20 men and 20 women—conduct 60-to-90-minute one-on-one in-depth interviews, and spend over six weeks on transcription, coding, and comprehensive analysis. The final report, distilled from 2,400 to 3,600 minutes of interviews, does capture meaningful patterns. Yet the client worries that purchase-history data from tens of millions of consumers nationwide may offer greater market representativeness for rapidly shifting preferences—and prove more useful for new-product development—than qualitative findings from only 40 informants.

This tension is structural to qualitative research and has long been a persistent problem. To address it, Kawakita [37, 38, 39, 40] developed the KJ method, in which multiple analysts start from 100 to 200 labeled data cards, extract meaningful groupings, and arrange them until the structure can be articulated. Glaser and Strauss [28] likewise proposed the Grounded Theory Approach (GTA)—extracting concepts and building up to higher-order "categories"—to reduce analyst bias. Neither method aimed at anything arbitrary; both maximally reflected the cognitive capacity of a skilled analyst working without computational support. Going back further, Peirce's [50] abductive reasoning formulated a third form of logic: inferentially generating the most plausible explanation from observed data.

In today's mature consumer markets, the premise of a "manageable volume of data" no longer holds. Digital touchpoints—URL and social-media access logs, reviews, customer-service records, online-community posts—together with chat-based interviews enabled by generative AI now generate qualitative data far surpassing past surveys, while practitioners must deliver insights ever faster. The "Global Market Research 2024" report by the European Society for Opinion and Marketing Research (ESOMAR) [22] likewise noted that marketing-research professionals face structural pressure to deliver richer insights from larger, more heterogeneous data. The scalability ceiling of traditional qualitative analysis has thus become a strategic bottleneck for embedding qualitative approaches in marketing.

Concerns about qualitative analysis have also been raised from another angle. In a large-scale replication project published in Nature, the independent replication rate across 164 papers from 54 social- and behavioral-science journals was only 55.1% [59]. Even more significant is the analytical-robustness study by Aczel, Szaszi et al. [1]: when multiple analysts re-analyzed the same data across 100 independent cases, complete agreement with the original papers was only 34%. This was read as pointing not to replicability but to the analyst itself being a source of bias. It can be seen as a structural phenomenon expressed numerically—the "differences in validation criteria across journal-based communities" noted by Fujigaki [24]. Because the scalability problem can amplify bias as analysts are added, it must be understood as a twofold crisis of scale and structural bias.

## 1.2 The Advent of LLMs: Possibilities and Remaining Challenges

The advance of large language models (LLMs) is accelerating qualitative analysis' processing capacity and automation. Pi et al. [51] reported 80–88% agreement with expert analysts' coding in Grounded Theory tasks, and Wen et al. [65], proposing "Neo-Grounded Theory (NGT)," report a 168-fold throughput gain at minimal cost. Übellacker [60] conducted an early proof of concept for automating a Grounded Theory workflow with GPT-4. Furthermore, prior work—CollabCoder by Gao et al. [25],

LLooM by Lam et al. [42], Auto-TA by Yi et al. [66], SFT-TA by Yi et al. [67], and Thematic-LM by Qiao et al. [53]—has extended this area toward human–AI collaborative coding, concept induction, fine-tuning agents, and multi-agent thematic analysis.

Through these prior studies, we raise five problems.

First, all existing systems rely on general-purpose LLMs that have not internalized domain knowledge through CPT on a domain-specific corpus. As Prescott et al. [52] showed, such models exhibit a substantial drop in agreement with human coding at the level of individual coding decisions, revealing limits in domain-specific interpretation.

Second, existing systems analyze only a single corpus and tend to overlook the cross-cutting analysis of qualitative data gathered from diverse sources—indispensable for extracting consumer insights.

Third, we have found no system that proposes evaluation metrics sensitive to the specific task of "consumer-insight extraction." Standard Natural Language Processing (NLP) metrics measure surface-level textual similarity rather than analytical value.

Fourth, existing systems end at the generation of codebooks or theme structures, leaving unachieved the connection to the business and marketing strategy formulation and tactical implementation that should follow insight extraction.

Fifth, the entire qualitative-research field those frames "consumer insights" rests on Western traditions—Grounded Theory and Thematic Analysis—and thus harbors the WEIRD (Western, Educated, Industrialized, Rich, Democratic) problem noted in psychology and the social sciences, failing to engage with non-Western methodologies.

### 1.3 Proposed Approach: Computational KJ-Hō (the Kawakita Jiro Method)

We propose "Computational KJ-Hō (the Kawakita Jiro method)." It is a theoretical framework that computationally realizes the KJ method's theoretical conviction—that the structuring of meaning should not be constrained by the analyst's pre-existing conceptual framework but should rest on the data (information) itself—so as to address the five problems above.

We call this orientation "analyst-bias-free," though the term requires careful definition. It does not mean the system operates free of any premises; every trained model inherits premises from its data. What is excluded are the analyst's conscious preconceptions—biases and assumptions from past experience, such as "What consumption patterns can we expect?" or "What consumer insights would be useful for this business?"—that steer results before the data are examined. Today's consumers change rapidly; recent advances in AI, in particular, are causing seismic shifts in consumer behavior. In this sense, the framework is more accurately "hypothesis-generating" than "hypothesis-free": rather than verifying a predefined structure, it generates explanatory structure from the data. A full operational definition is given in Section 4.4.

This framework is conceived as a three-layer architecture. Figure 1 presents the overall conceptual framework.

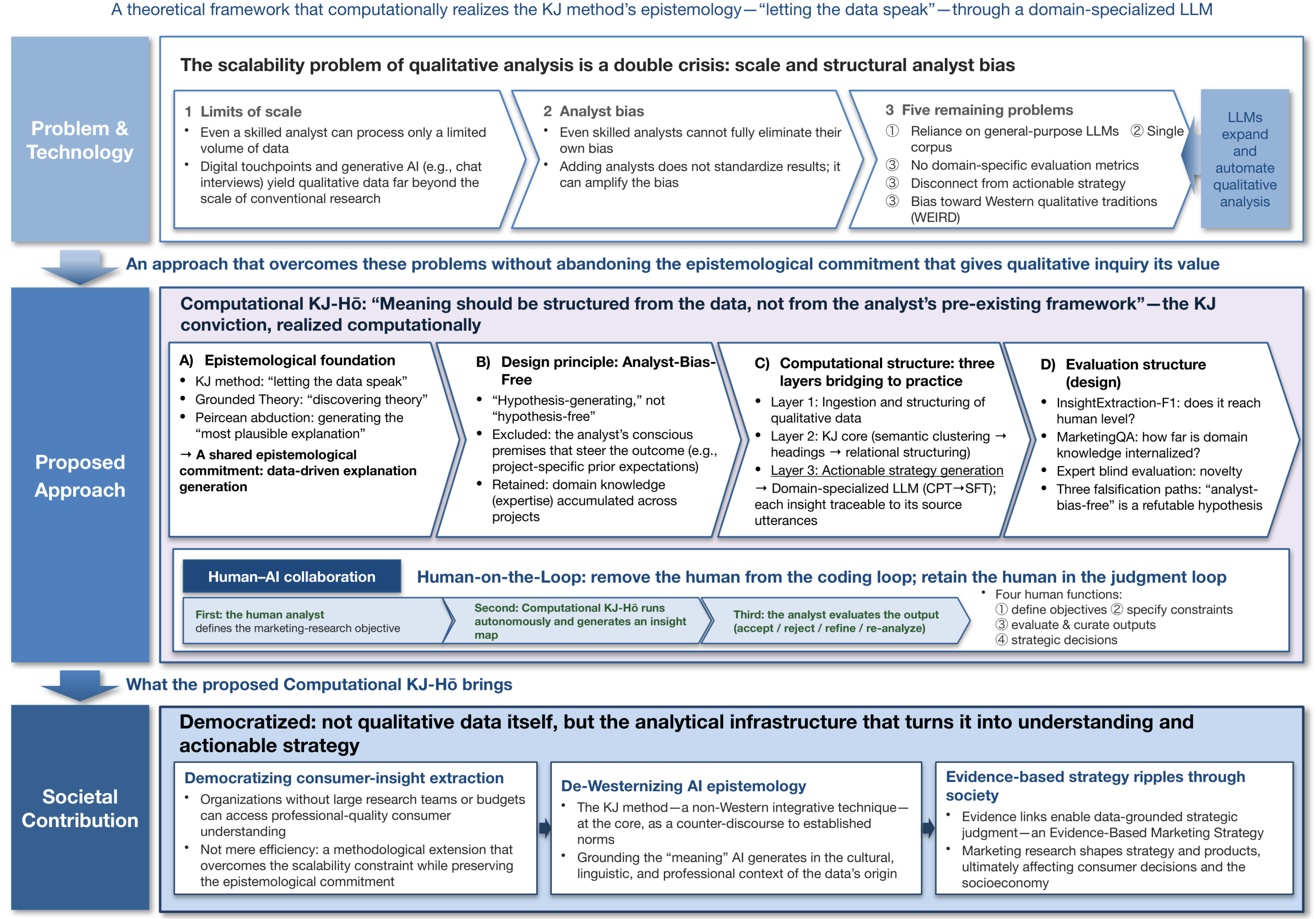

**Figure 1:** Conceptual framework of Computational KJ-Hō — problem and technology → proposed approach → societal contribution

The first layer handles the ingestion and structural preparation of qualitative data—here meaning text produced by consumers, such as in-depth interviews and online reviews.

The second layer is the computational re-creation of the KJ method's concept and the core of the architecture. It converts text into structured data and assigns labels, generates groups through clustering and related techniques, adjusts group levels, builds structural understanding such as mapping relationships among groups, and generates insights (as statements). All of this is handled consistently by a domain-specialized LLM (training procedures are described in Section 4).

The third layer converts the structured insights into executable marketing tactics, including target-segment definitions for business and marketing strategy, message concepts to communicate, and, where needed, proposals for quantitative-survey design. This extension from analytical output to strategic application means the framework aims at a practical contribution beyond existing academic and systems theory.

This study takes on social practice through advanced information processing in the qualitative-analysis domain that demands skilled analysts. Marketing research shapes firms' market understanding and consumer-insight acquisition, influencing strategy and product development, ultimately affecting consumer decisions and rippling through the socioeconomy; it is now used by public institutions as well. Hence the analyst bias latent in the process—especially WEIRD-centric epistemology imposed on non-WEIRD markets—is, in a society where diversity matters, a social and ethical problem beyond the quality of any single study. The "analyst-bias-free" orientation aims, through a domain-specialized LLM, at the structural reduction of such social bias. Moreover, the marketing research this framework supports is an inherently collaborative knowledge practice involving multiple analysts and stakeholders; we return to this positioning in Section 6.3.3.

### 1.4 Contributions

To restate, this study offers five contributions to the use of Human-Computer Interaction (HCI) in qualitative analysis.

First, it proposes a novel, scalable theoretical framework by integrating three intellectual traditions—the KJ method as the axis, Grounded Theory, and abductive reasoning—into a computational paradigm for "analyst-bias-free" consumer-insight extraction. To our knowledge, no prior study places a non-Western qualitative methodology (the KJ method) at its core.

Second, it presents a three-layer architecture that analyzes multiple consumer-generated qualitative datasets cross-sectionally. Most existing systems complete analysis within a single corpus (e.g., a single interview), making cross-dataset meaning extraction difficult. This study overcomes that with embeddings specialized for the marketing-research domain (an embedding is a representation of textual meaning as a numerical vector; hereafter, domain-specialized embeddings). The key technical feature is a two-stage training that goes beyond prompt-based correction: Continued Pre-Training (CPT) adapts the LLM itself to marketing research, and Supervised Fine-Tuning (SFT) equips it with the analytical operations the second layer requires.

Third, it proposes two newly designed metrics—InsightExtraction-F1 and MarketingQA—to evaluate insight quality in domain-specific contexts that standard NLP metrics cannot capture. These do not compete with Pi et al.'s [51] five codebook-quality metrics but complement them.

Fourth, it explicitly engages with the WEIRD problem [29]—long noted in psychology and the behavioral

sciences—holding that a continuity of the problem exists in data-driven qualitative analysis as well. Most existing systems were developed and validated in Western, English-language contexts. By placing a Japan-originated methodology at the core and planning cross-cultural validation in non-WEIRD markets, this study presents a standard that treats cultural validity as a design requirement rather than an afterthought.

Finally, a fifth contribution cuts across all the above. The framework is designed from the author's own practical experience of nearly three decades of marketing research not only in Japan but across North America, Europe, China, Southeast Asia, India, and Australia. The five problems in Section 1.2 are not mere literature observations; they are constraints encountered in practice that existing systems cannot resolve. The fifth contribution is not the experience itself, but the problem formulations and design requirements derived from it—the integration of domain expertise with computational methodology—which is, to our knowledge, currently unique in LLM-based qualitative-analysis research.

### 1.5 Positioning of This Paper

The ultimate destination of this study is a practical contribution—through the three-layer architecture above—that transcends existing academic and systems theory. Validation through collaboration with industry is indispensable and will require considerable effort and time, and the AI field develops at remarkable speed. Rather than letting the framework lie dormant until validation is complete, we have chosen to put it to the world before empirical validation and to seek opinions widely. This paper is therefore a concept paper announcing a theoretical framework, and a declaration of ongoing research.

## 2 Intellectual Foundations

### 2.1 The KJ Method [37]: Bottom-Up Integration as an Epistemological Commitment

The KJ method, originating from qualitative data, was developed by Jiro Kawakita, a Japanese geographer and cultural anthropologist; its name derives from his initials. Its details were first set out in his 1967 book Hassōhō [37] and refined through subsequent works, including Zoku Hassōhō (1970) [38], KJ-hō: Konton o shite katarashimeru (1986) [39], and a revised edition (2017) [40]. The method is designed for teams to organize and synthesize the diverse qualitative data accumulated through fieldwork records, interviews, and ethnography. Kawakita described its core process as three consecutive operations: "collecting information > organizing and structuring information > consensus building / collective decision-making." Specifically:

(1) Collect as much information as possible, fragment it, and create "labels" (one piece of information per card).
(2) Form groups by consolidating similar items and assign a "heading (nameplate)" to each group.
(3) Spatially arrange the resulting groups and structure (diagram) the relationships among them, using prescribed symbols such as "flow type (chronological/causal)," "satellite type (causal/interdependent)," "tree type (logical)," and "cycle type (cyclical)," while also reflecting semantic proximity in the spatial positioning.
(4) Finally, render the diagram into prose. Referring to each group's "heading," the team looks back together, confirms the diagram's validity, and revises it as needed—that is, reaching consensus on their analytical results and making a collective decision.

What distinguishes the KJ method from other qualitative-data techniques such as coding lies not only in its procedure but in its epistemological commitment. Its principle is that the structure among groups should rest on the data itself, emerging through iterative, intuitive integration, rather than being projected onto the data by the analyst's prior theoretical framework. Kawakita expressed this as "letting the data speak" [39]. This is a principle that researchers, when checking their own questions, doubts, or beliefs, must resist the temptation to let them control their judgment of the data. The KJ method means not starting from existing concepts, theories, preconceptions, wishful thinking, or hypotheses—that is, from "the common sense of the world the analyst lives in."

The KJ method is a "method of idea generation (Hassōhō)." Collecting qualitative information and data (= facts) and organizing them are merely means; the ultimate goal is for the researcher to generate new "ideas" through these means. This is why "consensus building / collective decision-making" matters. It is not a means for verifying what is already known.

This stance, however, entails a practical constraint. The cards handled in the KJ method have conventionally been said to number about one to two hundred—described as the optimal amount a single analyst can "physically" handle, the limit of what can be spread on a large table or the floor and taken in at a glance, and the number that human working memory can hold cognitively at once. So we have been taught—yet the author has always felt even that to be too many (playing the card game Concentration, could anyone remember the contents and positions of all 52 cards?). Beyond this threshold, the holistic, intuitive integration the method requires breaks down, and the analyst can no longer adequately perceive the relationships among data units. The human mind then drifts toward the easy path, and the emergent structure that should be discovered risks being driven by whichever labels happen to be salient or forced into a pre-existing mold.

This very constraint is the starting point of the Computational KJ-Hō framework. What the author questions is not whether the KJ method is epistemologically sound. Rather, having embraced data-emergent categorization as foundational, the question is whether—while preserving the method's epistemological character—we can computationally extend its handling of data volumes beyond an individual analyst's cognitive capacity. Furthermore, drawing on knowledge specialized to marketing research, the aim is a quantitative reproduction of the qualitative analysis entrusted to the skilled marketing researcher.

For readers unfamiliar with the KJ method, it is widely recognized as a structured methodology for organizational knowledge integration and is applied not only in Japan but in Japanese firms' overseas product development, public policy, and qualitative research; in Western-language scholarship, however, its treatment remains limited. The reference to it here is both a substantive methodological choice and one answer to a continuing problem in data-driven qualitative analysis—the WEIRD problem (engagement with non-Western intellectual traditions).

The principal theoretical challenge in computationally realizing the KJ method is whether the spatial arrangement based on Kawakita's "felt closeness" can be realized computationally. This judgment excludes preconceptions and values human intuition; this study interprets it as the analyst sensing proximity among cards through gestalt-like intuition—grasping the whole at once rather than assembling parts. It can be functionally approximated through domain-specialized vector similarity, because "felt closeness" is a judgment of semantic proximity whose aggregation becomes the very element of consumer insight. Domain-specialized CPT, learning from a large body of knowledge, places labels that experts would group together in proximate positions in the embedding space. This study does not, however, equate the KJ method's grouping with Hierarchical Agglomerative Clustering (HAC). HAC is a mechanical procedure based on a similarity matrix, different in character from the embodied, phenomenological affinity judgment the KJ method presupposes. A domain-specialized embedding model encodes the contextual knowledge that underpins what expert analysts "feel" to be close, and we hold that proximity relations in its embedding space can functionally approximate the semantic proximity on which KJ-method grouping relies. The two differ in internal mechanism, but whether the resulting groupings coincide is an empirically testable question. Section 4.3 reports preliminary evidence that the conceptual pipeline produces analytically coherent groupings; full-scale verification of the domain-specialization claim is carried out in the empirical phase, based on the evaluation design presented in Section 5.

### 2.2 Grounded Theory [28]: Hypothesis-Generating Analysis

The Grounded Theory Approach (GTA) is a theory advanced by sociologists Barney Glaser and Anselm L. Strauss in their 1967 The Discovery of Grounded Theory [28], deliberately contrasting with the then-dominant "hypothesis-testing" approach. It is a systematic methodology that extracts concepts—not summaries—from data and theorizes the mechanisms by which phenomena arise, following an analytical process that codes qualitative data and raises abstraction through classification. Although the founders later diverged and the approach has drawn much criticism, we value the "theory construction grounded in data" they originally aimed at, without relying on given analytical scaffolding. The original criterion of validity was "theoretical saturation"—the point at which further data collection and coding no longer generate new conceptual categories.

The two founders later diverged. Glaser [27] held that categories should avoid imposing any external conceptual structure and rest solely on the data. Strauss, with Corbin, adopted a more structured approach incorporating existing analytical frameworks to guide coding [15, 16]. The orientation toward data-grounded theory construction has also been carried on in constructivist grounded theory (Charmaz [12, 13]) and in systematic qualitative-analysis methodology for social scientists (Strauss [57]).

This study does not take the founders' dispute as its theme but stands closer to the Glaser lineage, because the "analyst-bias-free" orientation I propose minimizes the "hypothesis-testing" premise characteristic of marketing research and prioritizes emergence over imposition.

Grounded theory's relevance to this study is that it formally articulates what the KJ method embodies in practice: that qualitative analysis is essentially hypothesis-generating, not hypothesis-testing. Glaser and Strauss [28] argued that the goal of qualitative inquiry should be the discovery of theoretical explanations truly grounded in observed patterns, not the confirmation of propositions derived from prior theory. This matches what I call "analyst-bias-free": not a denial of all premises, but a systematic effort to minimize the degree to which the analyst's prior framework dictates what can be seen in the data.

### 2.3 Abduction [50]: The Logical Structure of Insight

The logical foundation common to both the KJ method and grounded theory traces to the abductive reasoning of the philosopher Peirce [50]. Peirce distinguished three forms of inference (Table 1): deduction, which applies a rule to a premise to derive a necessary conclusion; induction, which generalizes a universal rule from observed cases; and abduction, which selects or generates the most plausible hypothesis to explain an observed fact. Peirce regarded abduction as the only form of inference that produces genuinely new ideas: deduction and induction can confirm or corroborate existing knowledge, but only abduction can discover (Yonemori [68, 69]; Uchida [61]; Endo [19]).

Abduction follows a characteristic logical structure. Given a conclusion B and a rule "if A then B," one infers A as a premise. This is not arbitrary: it is constrained by the requirement that the hypothesis be the most plausible available explanation given the current state of knowledge. Abduction is therefore neither free invention nor mechanical extrapolation, but disciplined creative inference under uncertainty.

**Table 1: The Three Forms of Inference (tabulated by the author, based on Inoue [32])**

| Inference Method | Thought Process | Characteristics | Example |
|---|---|---|---|
| Deduction | General principle (premise) → specific conclusion | [Certainty] The principle must be true. It reaches a logically valid conclusion but produces no new ideas. | Major premise: All the beans in this bag are white. Minor premise: These beans were drawn from this bag. ↓ Conclusion: These beans are white. (specific conclusion) |
| Induction | Multiple observed results (facts) → general conclusion | [Generalization] Accumulating facts leads to the discovery of a general law. The conclusion can be overturned if an exception appears. | Minor premise: These beans were drawn from this bag. Conclusion: These beans are white. ↓ Major premise: All the beans in this bag are white. (generalization) |
| Abduction | Observed result (fact) → cause (hypothesis) | [Possibility] Infers the cause from the result. It is uncertain. It leads to new discoveries and the emergence of ideas. | Major premise: All the beans in this bag are white. Conclusion: These beans are white. ↓ Minor premise: These beans were drawn from this bag (hypothesis). |

This study holds that both the KJ method and grounded theory are, at their logical core, abductive processes. In each, the analyst observes a new regularity within facts that existing rules or patterns cannot reasonably explain, generates a hypothesis to make that regularity rationally intelligible (grouping, coding, categorization, theorization, and the like), and issues provisional results. The two differ in procedure—the KJ method visualizes and narrativizes the spatial-structural relationships among groups, while grounded theory raises abstraction through coding and classification—but both share the abductive commitment of generating explanation from observation rather than testing hypotheses.

With the introduction of a domain-specialized LLM, this study introduces the concept of "functional abduction." This is the force by which a domain-specialized model generates consumer insights that are absent from its training data and not specified as analytical hypotheses, and it characterizes this study. Its epistemological status is revisited in Section 6.1; for now, treat it as one element characterizing what the framework is designed to do.

### 2.4 Integration: Data-Driven Explanation Generation as a Unifying Principle

The three intellectual foundations surveyed above—the KJ method, GTA, and abduction—share a common epistemological commitment expressible as a single principle: that explanatory elements should be generated from the data rather than imposed on it from outside. Each tradition expresses this differently: Kawakita speaks of "letting the data speak" [39], Glaser and Strauss of "discovering theory" [28], and Peirce of generating the "most plausible explanation" [50] for observed facts. Despite differences in emphasis, they converge on the same underlying orientation toward empirical inquiry.

Computational KJ-Hō is designed as the computational realization of this shared principle. Taking "data-driven explanation generation"—that is, "data-driven hypothesis-generation"—as its foundational design criterion, it asks how a domain-specialized LLM can be deployed to extend this commitment to data scales and analytical contexts beyond human cognitive capacity. Table 2 compares the three traditions across four analytical dimensions.

**Table 2: Comparative Overview of the Three Traditional Intellectual Foundations**

| | **KJ Method** | **Grounded Theory** | **Abduction** |
|---|---|---|---|
| Origin | Developed by Jiro Kawakita (Japanese geographer and cultural anthropologist) as a technique for organizing large-scale fieldwork; refined in a series of works beginning with Hassōhō (1967). | Proposed by Barney Glaser and Anselm Strauss (American sociologists) in The Discovery of Grounded Theory (1967), as a deliberate counter to the then-dominant hypothesis-testing approach. | Proposed by Peirce (American philosopher of pragmatism) from the late 19th to the early 20th century as a systematization of the forms of inference. |
| Core Process | ① Labeling (fragmenting information) → ② group formation and heading assignment → ③ diagramming (spatial arrangement and relational structuring) → ④ narrativization (prose writing and collective decision-making). | Open coding → classification of coded data → theorization that raises the level of abstraction. Theoretical saturation serves as the criterion of validity. | Generating the most plausible explanatory hypothesis from observed facts and existing rules. |
| Underlying Belief | "Let the data speak"—excluding the projection of the analyst's prior framework, preconceptions, and existing hypotheses, and respecting the structure that emerges from the data. | Hypothesis generation rather than hypothesis testing; aims at discovering theoretical explanations genuinely grounded in observed patterns. | "Disciplined creative inference under uncertainty"—the only form of inference that produces genuinely new ideas (neither arbitrary nor mechanical, but the most plausible explanation given the current state of knowledge). |
| Issues in Application | About 100–200 cards (the upper limit a single analyst can physically and cognitively handle). Beyond this threshold, holistic, intuitive integration breaks down, risking being forced onto salient labels or pre-existing molds rather than emergent structure. | Depends on the analyst's theoretical sensitivity. The founders' divergence (Glaser vs. Strauss–Corbin) has produced methodological indeterminacy. | A philosophical logical form; concrete implementation procedures are not specified. |

The convergence of these three foundations suggests that "data-driven hypothesis-generation" is not the idiosyncratic feature of any single methodology but a widely recognized epistemological desideratum in qualitative inquiry.

Computational KJ-Hō adapts this desideratum to the analysis of large-scale qualitative data through the following design principles:

(1) A domain-specialized embedding model that reflects the semantic structure of the marketing-research domain for understanding consumer behavior, rather than general linguistic patterns.
(2) A bottom-up approach that generates groupings from the affinities in the data, rather than from a predefined system of categories or groups.

(3) Emergent structure visualization that renders the structural relationships emerging from the analyzed data explicit and explorable.
(4) “Insight verbalization” whose goal is to generate human-readable explanatory descriptions grounded in concrete evidence.

By starting from these traditional foundations, this study is not merely the development of a technical system for processing qualitative data more efficiently, but an inquiry deeply rooted in multiple scholarly traditions, aiming to establish a coherent methodological approach. The environments and mechanisms that realize the computation serve the epistemological commitment; they do not replace the inquiry itself.

# 3 Related Work: LLMs in Qualitative Research

## 3.1 Full Automation and Semantic Resolution

In recent years, many studies have applied LLMs to automating qualitative analysis. This section surveys the main prior work in order of proximity to Computational KJ-Hō, focusing on differences in "semantic resolution."

LOGOS [51] and Auto-TA [66] are important prior studies that extend traditional foundations to LLMs and automate the entire pipeline. LOGOS built a hierarchical codebook via graph reasoning. Auto-TA automated Steps 1–3 of Braun & Clarke's [10] Reflexive Thematic Analysis (RTA) using four role-conditioned agents and an iterative refinement loop, reporting high agreement with experts (up to 88% for LOGOS). However, relying on general-purpose LLMs (e.g., GPT-4) without internalizing domain knowledge through CPT leaves room for improvement.

Neo-Grounded Theory (NGT) [65] by Wen et al., using high-dimensional vector clustering and a multi-agent system, advanced the automation of grounded theory and achieved a 168-fold speedup over manual analysis; in terms of technical composition, it is the closest prior work to this study. The difference lies in domain-adaptation strategy: whereas NGT leveraged general-purpose embeddings (OpenAI's text-embedding-3-small), this study follows the BioBERT [43] paradigm and, through CPT, reshapes the embedding space itself into the symbolic system of marketing research—aiming to improve resolution for distinctions often conflated in general spaces (e.g., "price sensitivity" vs. "perceived quality," "high price" vs. "premium feel").

The two approaches are complementary, each useful for the analytical quality domain specialization requires and the speed general models pursue. Wen et al.'s implementation used agglomerative (hierarchical) clustering on OpenAI text-embedding-3-small (1,536 dimensions), achieving 0.904 on the quality score defined in that paper (manual coding: 0.883) on 40,000 characters of Chinese interview data. This study differs in design philosophy—reshaping the embedding space itself via CPT rather than selecting thresholds in a general-purpose embedding space—but we set this result as a target.

AcademiaOS [60], LLooM [42], and HICode [70] explored the automation of inductive coding through different approaches. LLooM identified as a problem that the low-level keyword extraction into which existing topic models (such as LDA and BERTopic) fall diverges from human semantic understanding and judgment and proposed and implemented extraction at a higher conceptual level. HICode applied a two-stage pipeline (label generation plus hierarchical clustering) to inductive coding. Neither adopts CPT to reshape a domain-specific semantic space, and thus they differ from what this study aims for. Thematic-LM [53] reduced hallucination by assigning "identity perspectives" through a multi-agent design; but this deliberately injects diverse identities to "pluralize bias," an orientation different from this studies.

The limits of general-purpose applications built on the standard Thematic Analysis (TA) framework [10] are also becoming clear. Naeem et al. [46] sought to address TA's long-criticized subjectivity, analyst bias, and limited capacity for large-scale data by embedding general-purpose LLM prompts (ChatGPT) into its six steps. Although AI improved comprehensiveness, researcher supervision and judgment remained indispensable at each step, leading to a division-of-labor model. Ayik et al. [6] comparatively evaluated the six TA steps across four tools (ChatGPT-4o, QInsights, ATLAS.ti AI, MAXQDA AI Assist), finding that AI contributes time savings but does not fully reproduce a skilled marketing researcher's interpretive depth. These findings reinforce this study's claim that bridging the gap requires internalizing domain expertise through CPT.

SFT-TA [67], by the same research team as the aforementioned Auto-TA [66], proposed an inductive-TA automation framework embedding Supervised Fine-Tuning (SFT) into a multi-agent system. A pipeline embedding SFT agents in both the coding and theme-generation phases achieved the highest performance. In terms of training strategy, this is the prior work this study should most closely reference. Computational KJ-Hō rests on two-stage training—CPT with domain-specialized embeddings to adapt the LLM itself to marketing research, then SFT to acquire the required analytical operations—ensuring that domain-knowledge internalization precedes task adaptation and aiming for a more rigorous "functional abduction."

### 3.2 Human–AI Collaborative Approaches

Collaborative Qualitative Analysis (CQA), in which multiple analysts or teams cooperate to analyze qualitative data, contributes to reducing bias, deepening interpretation, and improving the reliability of analytical results. However, it requires considerable time and effort, and it faces a dilemma: the more researcher independence and diversity are secured, the harder consensus becomes to reach.

CollabCoder [25] proposed an end-to-end workflow integrating an LLM into the three main stages of CQA (Independent Open Coding, Merge and Discussion, and Code Group Generation), implementing—in a study with 16 users—on-demand code suggestions, similarity scores and GPT-mediated proposals toward final code decisions, and GPT-based automatic generation of codes from agreed-upon codes. The design positions the LLM as an interactive partner that augments rather than replaces human judgment—an orientation that resonates with CQA—but it does not resolve the “scalability ceiling of traditional qualitative analysis methods” that I regard as the strategic bottleneck to embedding qualitative approaches in marketing.

ThemeViz [34] combines GPT-4 with interactive visualization to support the theme-development stage (Phases 3–5 of TA). Against AI-assisted QDA research that has traditionally focused on coding, its novelty lies in examining whether an LLM can serve as a collaborator in the more abstract, conceptual theme development. Its empirical evaluation showed greater utility than ChatGPT and exhibited no observed hallucinations; yet 27 of 28 participants regarded the AI as a “tool” rather than an equal collaborator.

MindCoder [26] (centered at Johns Hopkins) maps the entire qualitative-analysis process onto the codes-to-theory model and automates it as a Chain-of-Thought reasoning chain. Its core is trustworthiness: mechanical work (grouping, validation) is delegated to the system while interpretation remains human, and complete logs make the LLM's reasoning auditable. Evaluation by 12 users and 2 external reviewers confirmed active interpretive support and more trustworthy codebooks. Challenges remain: effectiveness varies with data type and scale, and verification used only GPT, requiring validation of generalizability.

Prescott et al. [52] compared, for both inductive and deductive TA, theme consistency, reliability, and working time between GenAI (ChatGPT-3.5, Bard) and human coders, analyzing 40 SMS messages from a digital intervention promoting medication adherence among patients with HIV. Theme-level consistency was acceptable, but the reliability of individual coding decisions was substantially lower on the AI side, which also showed limited understanding of context and subtle meaning—such as slang related to drug use—and could not reproduce the themes humans found. Prescott et al. therefore concluded that a hybrid (division-of-labor) approach is appropriate. This is strong motivation for this study's CPT approach: the solution lies not in compensating through prompting but in reshaping knowledge representation through domain-specific pre-training.

### 3.3 LLMs in Marketing Research

A rapidly expanding body of work has applied LLMs to marketing-research tasks in fundamentally different modes.

#### *3.3.1 LLMs as Synthetic (Surrogate) Respondents*

Brand et al. [8, 9] showed that LLM responses to conjoint surveys are statistically comparable to those of human consumers, and that fine-tuning on prior survey data further improves this alignment. This line positions LLMs as synthetic respondents—a computational substitute for human consumers in survey-based research.

This is fundamentally different from Computational KJ-Hō. Whereas Brand et al. and Arora et al. [4] use LLMs to simulate consumers' behavior and mindset, this study uses LLMs to analyze what consumers actually said. The former addresses the cost and speed of data collection; this study addresses the depth and scale of data analysis. Research on "AI consumers" (fictitious synthetic consumers) and this study differ in purpose, target, and data, so they do not compete—indeed, they are complementary in the evolution of marketing research.

Regarding the possibility of LLMs substituting for marketing research, affirmative, critical, and practitioner positions run in parallel.

#### *3.3.2 LLMs as a Substitute for Quantitative Surveys*

As noted above, Brand et al. [8, 9] showed that LLM-generated Willingness-to-Pay (WTP) estimates in conjoint analysis agree with humans at 64–90% accuracy, and Arora et al. [4] demonstrated that a hybrid design of LLM plus human analysts surpasses humans alone in insight into qualitative data. Asirvatham et al. [5] proposed the Generalized Attribute-Based Ratings Information Extraction Library (GABRIEL), a Python library that scores text, image, and audio on a 0–100 scale against natural-language-defined attributes, reporting that an LLM (GPT-5 family) achieved statistical accuracy indistinguishable from human annotators, with projected cost reductions of 700-fold (GPT-5) and 17,500-fold (GPT-5-nano) versus a human annotator (at $15/hour). The utility of generative AI in knowledge generation has also been reported (Estevez et al. [23]).

On the other hand, Bisbee et al. [7] showed that 48% of the regression coefficients from LLM responses differed significantly from human data, of which 32% had reversed signs, and that results varied over time even for identical prompts—non-reproducibility. Dutt [54], moreover, warned of LLMs' limits in emotion- and context-dependent judgments and of the risk of sycophancy.

In research on simulation infrastructure, Wang, Z. et al.'s [63] OPeRA dataset showed that even the best-performing model, GPT-4.1, reached only 22% generation accuracy for human behavior, while Ren et al.'s [55] BASES achieved a 13% relative improvement in NDCG@1 via two-stage prompting. Overall, however, the field converges on the view that full substitution by LLMs alone is currently impossible and that hybrid designs are optimal.

This study (Computational KJ-Hō) sits at a different problem layer. Because it automates the analysis of qualitative data from real humans rather than generating fictitious "respondents," the problems Bisbee et al. noted—failure to reproduce response diversity, non-reproducibility, WEIRD bias, and sign reversal—lie outside its scope. Rather, this study corrects the "epistemic limits of LLMs" those critiques identified through a dual structure: reshaping the semantic space via domain-specialized CPT, and human-expert evaluation of outputs.

### 3.4 Direct Extraction of Customer Needs: The Prior Work Closest to This Study in Terms of Task

Timoshenko et al. [58] reported that, in extracting customer needs from Voice of the Customer (VOC) data, an SFT-trained LLM matched VOC experts' accuracy with as few as 600 training examples. Notably, this small number of training examples yielded sufficient performance, with low dependence on the base foundation model and no observed hallucinations. The difference from this study is domain-adaptation strategy: whereas Timoshenko et al. adapted through SFT alone, this study places the reshaping of the semantic space via CPT at the foundation and then combines SFT. This guarantees by design that the internalization of domain knowledge precedes task adaptation, aiming for a more rigorous "functional abduction."

### 3.5 LLM-Based Analytical Support Toolkits: Prior Work Sharing This Study's Problem Awareness

DeTAILS [56] is a self-contained, cross-platform desktop application designed for large-scale qualitative data analysis. Based on the six-step TA proposed by Braun & Clarke [10], it implements an iterative workflow in a fully local environment, shuttling the codebook back and forth between researcher and LLM as a "boundary object"—an artifact that parties in different positions share as a point of collaboration. Notably, it guards against code hallucination by checking quotations against the source text and discarding mismatches, a design intended to preserve analytic agency. In an evaluation by 18 qualitative researchers (six each of novice, mid-level, and expert), analysis sped up, with no significant differences in F1 or other key metrics across researchers. On the other hand, the study's evaluation and discussion point to the following limitations.

(1) Diminished reflexivity: confirmation of AI output dominates, making deep reflexivity on one's own analytical premises less likely.
(2) Agreement vs. interpretive validity: high agreement between LLM and human codes does not guarantee meaningful interpretation.
(3) Novices' over-trust in AI: the more fluent the output, the more it is accepted without critical scrutiny.
(4) Limited backtracking: changes propagate only to later steps, with no automatic correction back through the process.

These findings empirically reinforce this study's position that prompt-based correction of general-purpose LLMs alone cannot reach "analyst-bias-free," and that a dual structure—domain-knowledge internalization via CPT and human-expert evaluation of outputs—is essential. Moreover, whereas DeTAILS relies on RAG for external knowledge, this study differs in design by placing the reshaping of the semantic space via CPT at the foundation of analytical quality.

### 3.6 Methodological Reflection

Ibrahim and Voyer [31] pointed to "Technological Reflexivity" when deploying LLM-based tools in qualitative research—extending "researcher reflexivity" to AI as six elements of self-reflection researchers should perform. They hold that, because LLMs are made of text and sensitive to social meaning, they are more congenial to the qualitative paradigm—which values subjectivity and interpretation—than to the quantitative paradigm, which emphasizes statistical validity and generalizability. This resonates with this study's use of LLMs as a qualitative, interpretive approach. Accordingly, Section 4.4 discusses what excluding analyst bias means, and Section 6.3 the requirements for human oversight and process transparency.

Through the body of prior work surveyed in this chapter, the following five common problems are reconfirmed:

(1) Reliance on general-purpose LLMs without CPT-based domain adaptation.
(2) Confinement to a single dataset, lacking a cross-cutting segment-pattern perspective.
(3) The absence of evaluation metrics tailored to the domain.
(4) Analysis is treated as an end in itself, lacking a perspective that encompasses the entire business through to the strategic output that should ultimately be considered (i.e., marketing decision-making).
(5) And resting solely on Western qualitative-research traditions.

Computational KJ-Hō is designed to address these five problems, as shown through the concrete claims of the "computational structure" and "evaluation structure" in Section 4 onward.

# 4 The KJ-Method Computational Structure

## 4.1 Overview of the Computational Structure

Computational KJ-Hō is composed of a three-layer architecture, with each layer corresponding to a distinct analytical function.

Layer 1: Ingestion and structuring of qualitative (raw) data

Layer 2: Extraction and structuring of consumer insights from that data

Layer 3: Connecting the structured consumer insights to marketing strategy and converting them into implementable output

The layers are designed sequentially, each layer's output feeding the next. Yet the analytical process is separable: each layer has its own design decisions and ensures output quality through independent evaluation criteria, enabling improvement.

Throughout, the structure is driven by a domain-specialized LLM built through a two-stage learning process.

The first stage performs CPT on a large-scale corpus related to marketing and marketing research. At minimum, the corpus requires: (1) information related to consumer psychology and consumer behavior (consumer-research reports, articles, marketing trade journals, specialized books, information from academic societies and relevant ministries); (2) information on business, economics, and law (securities reports, IR materials, stock information, information on relevant laws and regulations); and (3) academic literature (consumer psychology, brand strategy, consumer-behavior theory, social psychology, economics, and management). Also indispensable are (4) knowledge of research methods (quantitative and qualitative, across fields) and the latest research presented at academic societies and associations, and

(5) practical knowledge and academic achievements in marketing research—the accumulated wisdom of the skilled marketing researchers who have long elevated consumers' words into "consumer insights" informed by their living environments and deep psychology.

On top of this, we add this study's originality: the incorporation of "best practice" and "benchmarking" thinking. As Newton wrote of "standing on the shoulders of Giants" [47], practical knowledge learned from superior cases at other firms matters in practice, and we reflect it in the LLM. Specifically, (6) the company histories (shashi) of various firms serve as a corpus of benchmark information on business models and marketing strategy. Here, "shashi" denotes the "Japanese-style shashi" a firm compile on its own responsibility, covering its history, achievements, and internal processes—unlike the overseas type written by outside scholars or journalists. It is itself the primary data an analyst should value. Because using only enduring firms' histories would create "survivorship bias," bankrupt firms' histories are also included. These corporate corpora, containing best and worst practices alike, are important for the learning of the third layer.

The second stage performs SFT on expert-curated paired data. The preferred combination is "consumer-generated text (e.g., interviews) → consumer insights derived from them → the marketing strategy and measures built." This trains the model to perform the specific analytical operations the second layer requires.

That CPT precedes SFT is by design. Internalizing domain knowledge is a precondition for effective task adaptation. A model fine-tuned without first acquiring domain knowledge would end up applying analysis in a semantic space that does not adequately represent domain-relevant distinctions. This is the limitation Prescott et al. [52] noted. The CPT corpus comprises six text resources reflecting the semantic structure of the marketing domain; its exact composition will be fixed through curation before final training and reported in a subsequent evaluation paper. The SFT stage uses expert-selected paired data, adopting a quality-control protocol in which each sample passes mutual review by at least two independent curators.

The validity of this quality control is also supported by the findings of Alt et al. [3] in relation extraction. They noted that the quality of the most widely used crowdsourced dataset for Relation Extraction (RE) in Natural Language Processing (NLP)—the TAC Relation Extraction Dataset (TACRED)—was overestimated, and performed expert re-annotation. As a result, over 50% of the difficult data required re-labeling, and expert re-annotation improved the baseline average F1 from 62.1 to 70.1. They further showed that, under conditions where the relations are not substituted, models tend to adopt surface-level heuristics. This supports, with empirical evidence from another domain, this study's design choice to entrust SFT training data to mutual review by multiple expert curators rather than to crowdsourcing.

The model scale envisioned for empirical study is 17 to 27 billion parameters, premised on operation on local computing infrastructure without dependence on external APIs. This scale reflects a deliberate balance between analytical capability and operational autonomy. Adapting to the higher-order business domain of "marketing strategy construction" requires confidentiality management not only of input data but also of the analytical process and outputs. The model must therefore possess capability sufficient for complex analytical reasoning while being deployable—from fine-tuning to tactical prototyping—within a single research institution's or firm's computing environment.

## 4.2 The Three-Layer Architecture

### Layer 1: Ingestion and Structuring of Qualitative Data

The first layer handles the ingestion and structuring of qualitative data. For large volumes of qualitative data, it integrates speakers' attributes, extracts intent, emotion, and behavioral motivation from utterances, builds multidimensional representations linked to respondent attributes, and segments the data into individual utterances, exchanges, or thematically coherent fragments. These become the atomic data units of Computational KJ-Hō. Segmentation is performed by the domain-specialized model, which through CPT has acquired an understanding of the data's meaningful components; each utterance is given a domain-specialized embedding and located in the model's learned semantic space.

**Layer 2: Consumer-Insight Extraction**

The second layer automatically extracts consumer insights from the structured qualitative data without requiring hypotheses. It analyzes data cross-sectionally, detecting latent themes, extracting inter-segment differences, structuring consumer psychology, and identifying purchase motivations and barriers, then verbalizes and outputs insights with marketing significance.

Technically, the Computational KJ-Hō core process executes the KJ-method process of "grouping labels → assigning headings → spatial arrangement."

**Step 1—Semantic Clustering (KJ method: group formation)**

For the data structured in the first layer, whereas a general embedding model locates utterances relative to the full distribution of general language use, this study's post-CPT domain-specialized embedding locates them relative to the semantic structure of consumer-behavior utterances. Using embeddings from this model—which has internalized marketing-research vocabulary and consumer-psychology concepts through CPT—yields groupings aligned with consumers' latent motivations that general models miss. Rather than humans setting categories or code systems in advance, groups form bottom-up from semantic proximity in the data. Cross-tabulation with respondent attributes simultaneously detects patterns that differ across segments.

A practical consequence is the ability to distinguish utterances expressing conceptually close consumer attitudes. For example, in general embeddings "high price" and "premium" are close, but in consumer utterances the former often signals "dissatisfaction / overpricing" and the latter "aspiration / value recognition"—opposite motivations. The CPT-trained model internalizes such semantic shifts and yields grouping structures more directly tied to action design than feeding general embeddings into BERTopic and the like.

The domain-specialized embedding model, without referring to a predefined category system, generates clusters of semantically related data via methods such as hierarchical agglomerative clustering. The number of clusters is determined emergently from the data's embedding structure through a stopping criterion tuned in SFT, preserving the KJ method's commitment to data-driven grouping. Crucially, clustering operates on the entire corpus simultaneously: a single cluster can draw from multiple interviews, segments, or time points, enabling the "ideation" of patterns undetectable in single-corpus analysis.

The computational realization of "closeness" through domain-specialized vector similarity seeks to maintain phenomenological replication. Domain-specialized embeddings encode the contextual knowledge underlying marketing-research experts' closeness judgments, placing data that experts would group together in proximate positions in the embedding space. The key mechanism is that CPT on a marketing-research corpus builds semantic representations that structurally encode distinctions relevant to consumer psychology—for example, price sensitivity versus perceived value—approximating the knowledge that grounds experts' grouping intuitions. The philosophical basis was detailed in Section 2.1. This finer semantic resolution is the foundation on which the second layer's clustering relies.

**Step 2—Assigning Meaning to Groups (creating headings)**

For each cluster, the SFT-trained model generates a natural-language insight (a "heading") that integrates the meaning of the constituent data into a marketing-significant proposition. Each generated insight carries an explicit evidence link to its supporting data, allowing human reviewers to audit the reasoning and verify the empirical basis. Unlike a general-purpose LLM that depends on the prompt designer's hypotheses and framing, the SFT-trained model—through training on expert-curated paired data—internalizes the expressions that a professional marketing researcher would recognize as analytically valuable.

**Step 3—Structuring the Relationships Among Groups (spatial arrangement and structuring)**

For the groups generated in Step 2, this step uncovers relationships—causal, contradictory, hierarchical, complementary, and the like—and produces a structured map that makes their relational configuration explicit and explorable. By making relationships explicit—such as the opposition between "purchase motivation A" and "barrier B," or that "latent need C" is a superordinate concept of "behavioral pattern D"—it prepares an actionable input for the strategy construction and planning of the third layer.

**Layer 3: Strategy Generation**

The third layer connects the insights structured by the second layer to marketing strategy and converts them into implementable output. Concrete output should include the following.

(1) A statement expressing the content and relational structure of the consumer insights (the verbalization of the structured map).
(2) The definition of a target (segment) carrying consumer insights derived from the cross-cutting cluster structuring.
(3) The definition of opportunities linking consumer insights to the business (brand) side under consideration.
(4) Proposed brand communication (message and creative ideas) to appeal to the target.
(5) And, naturally, the automatic generation of the analytical report underpinning items (1)-(4) (executive summary, detailed analysis, with cited evidence), together with proposals for additional research on issues not resolved in the present analysis.

The third layer does not directly correspond to the KJ method's core process (collecting > organizing/structuring > consensus building / collective decision-making); rather, it addresses the fourth problem in Section 1.2, extending the framework from academic methodology to practical business application.

### 4.3 The Role of Domain Specialization: Why General-Purpose LLMs Are Insufficient

The need for domain specialization is not merely a theoretical proposition. Preliminary studies [33, 48] using real consumer-generated text and general-purpose LLMs confirm that the analytical pipeline works at the conceptual level but does not reach the semantic distinctions that skilled marketing researchers consider important.

Ishimoto [33] compared automatic labeling by general-purpose models (using 7W2H and the ABC model of consumer behavior) against manual gold-standard labeling on 779 consumer reviews of sake (after preprocessing). Processing time was far shorter—about 10 minutes for the LLM versus 19 hours by hand—but agreement with human labels remained low: 0.091 for GPT-3.5-turbo and only 0.317 even for GPT-4o-mini. Explicit elements such as sentiment (0.783) could be captured, but fine-grained situational distinctions such as drinking temperature and price range fell far short of human performance. Moreover, whereas GPT-3.5-turbo produced an uninterpretable cluster structure, GPT-4o-mini reproduced the same three clusters and the same top-level classification structure as the human analysis (special-occasion oriented / diverse sweet / cost-performance oriented). The following year, Oyamada [48] conducted a follow-up study using 958 interview records actually collected via AI chat interviews (＊1) and GPT-5-mini, structuring the data with intermediate labels that operationalized the consumer-behavior framework "7W2H" (Who, What, When, Where, Why, Whom, Compared to what, How, How much/How many). Using multiple correspondence analysis and co-occurrence analysis, the study showed that consumption contexts can be extracted quantitatively as linkages of "situation × motivation × behavior" (a top combination example: mood change/reward × emotional satisfaction × relaxation consumption).

As preliminary studies, the two fulfill conceptual and methodological functions for this study and also help establish performance baselines for measuring the improvement from domain specialization presented in Section 5. Ishimoto's results show that a model-generation update alone lifts agreement substantially, from 0.091 to 0.317—yet even a capable general-purpose model does not reach the semantic precision domain experts require. Oyamada's study showed the effectiveness of understanding the context of consumption behavior—"what is consumed, in what situation, with what emotional motivation, and how"—rather than simply segmenting consumers. Note here that a framework like the 7W2H Oyamada used is not a prior hypothesis that steers the results of a particular project, but a domain-universal lens for describing consumption behavior. What this study excludes is the former (the analyst's prior hypotheses); the latter is domain knowledge that should be internalized in the model itself through CPT rather than supplied externally through prompts each time (Section 4.4).

The need for CPT-based domain specialization can also be shown through the following three complementary points.

1. Semantic space: domain-specific CPT reshapes the embedding space so that words generally treated as synonyms are reflected where their semantic differences matter within a specific domain. Lee et al. [43] reported that BioBERT outperformed general-purpose BERT on biomedical named-entity recognition and relation extraction, attributing the gain to domain adaptation through pre-training. Kawakami et al. [36] reported a similar finding for a medical LLM developed via CPT on a clinical corpus. In marketing research, too, consumer language is rich in subtle distinctions—e.g., between stated "preference intent" and revealed "preference behavior," between price evaluation and value perception, and between satisfaction and brand loyalty—distinctions that general-purpose models, for which such fine differences carry no special weight, do not reliably encode.
2. Internalized expertise: SFT on consumer insights curated by skilled marketing researchers internalizes what counts as valuable analytical output in the context of marketing research. The model learns not merely to summarize consumers' qualitative data as clusters, but to generate the kind of synthesis an experienced marketing researcher would recognize as a consumer insight—evidence-based, strategically actionable, and ideally non-obvious. Humans provide only the research objective (and operational constraints); the analytical framework is supplied by the model itself. This contrasts with prompt-based approaches in which the framework is specified by the prompt designer's premises and assumptions.

3. Data sovereignty and cost: This is no denial of the continuing evolution of general-purpose models, but the primary reason for insisting on a domain-specialized model is sovereignty over data. The entire process by which a firm builds its management strategy, and the information involved, are normally confidential: the markets under consideration, consumer data, and hypotheses cannot leave the company. As a result, cloud environments—and tools premised on the cloud—in which the firm cannot control information flow are likely to be judged unadoptable. The second reason is cost. A self-hosted model in the 17-to-27-billion-parameter range can operate at roughly one-tenth the cost of a frontier commercial API on comparable tasks while keeping confidential data within a controlled environment. This is a structural argument grounded in the difference between usage-based API billing and the marginal-cost structure of self-hosting (once the GPU investment is made, per-inference cost is low), whose advantage grows with scale and continuous operation. A quantitative comparison will be reported in a subsequent empirical paper.

It is also worth examining the hypothesis that this study's effect could be approximated by alternatives without CPT: specifically, (a) domain steering of a general-purpose LLM via Few-shot Prompting, (b) reference to external domain knowledge via Retrieval-Augmented Generation (RAG), and (c) control of the analytical process via Prompt Engineering. The differences from this study are as follows.
(a) Few-shot remains surface-level pattern matching and does not reshape the embedding space; Webson & Pavlick [64] noted that GPT-3's [11] few-shot lacks sensitivity to template meaning.
(b) RAG can reference external knowledge, but because the internal semantic representation is not domain-adapted, meaning flattens when interpreting retrieved results.
(c) Prompt Engineering brings in the analyst's prior hypotheses and so does not reach "analyst-bias-free." These may hold cost advantages, but they cannot fully substitute for the depth, reproducibility, and bias reduction that CPT aims for.

The systematic positioning of these alternatives can be organized via Ling et al.'s [44] comprehensive survey of domain-specialization techniques (ACM Computing Surveys). Based on access to the LLM's parameters (black-box / grey-box / white-box), it systematizes domain-specialization methods into external augmentation, prompt crafting, and model fine-tuning. The alternatives (a)(b)(c) above correspond to the first two, while this study's CPT+SFT belongs to model fine-tuning (white-box). This study holds that domain-knowledge internalization is a precondition for analytical quality.

### 4.4 The Operational Definition of “Analyst-Bias-Free”

The advocacy of "analyst-bias-free" is an ambitious turn of phrase, and also a branding choice for a concept paper; in practice it should be understood as a methodology that structurally reduces analyst bias. It does not claim complete freedom from bias but, per the operational definition, a design that minimizes the intervention of the analyst's prior hypotheses in the early stage of analysis. Traditionally, the KJ method asks researchers to suspend subjective judgment and "let the data speak." This study realizes that computationally by placing a high-dimensional embedding space as the early-stage analytical foundation, where the semantic understanding and patterning of qualitative data emerge bottom-up from mathematical proximity, not from premises or hypotheses the analyst set in advance.

This approach does not claim quantitative or statistical objectivity. Rather, it embodies the "Technological Reflexivity" of Ibrahim and Voyer [31]. By using a CPT-refined model as a "Computational Mirror"—a mirror reflecting back collective domain knowledge—the system reflects not a single analyst's limited experience, but the collective domain knowledge distilled from many tokens. This also helps mitigate "thematic foreclosure"—the premature closing-off of themes in which researchers filter data early to fit existing frameworks.

This study assigns the LLM the role of proposing qualitative analytical results with analyst bias removed; within the framework discussed at the CHI workshop, this position corresponds to the left-hand side [21]. This does not replace human expertise but provides a rigorous starting point for final interpretation, strengthening the researcher's analytical accountability. By bridging the mathematical rigor of a specialized LLM and the holistic integration of the KJ method, it realizes a transparent, reproducible path from raw qualitative data to strategic consumer insights.

What this study insists on excluding is the analyst's rigidified thinking, bound to industry common sense and precedent. Anyone in marketing must be humble before the consumer and the qualitative data obtained from them. If bound by one's own "shoulds," the analyst's perception is impaired by those shackles. This differs from the usually recommended "hypothesis-driven" thinking. We do not deny hypotheses, but the diversity and speed of change of today's consumers exceed our assumptions; it is precisely when facing raw voices that the "shoulds" must be cast off. With a fine hypothesis, coding becomes easy—but data that contradict it are excluded and buried. Whether that is a correct understanding, no one verifies—the more skilled the analyst, the less so. What "analyst-bias-free" targets is precisely this skilled person's fixation—and pride. A hypothesis is only a hypothesis. We should, for a moment, become "hypothesis-free" and adopt the posture of letting the data speak—now more than ever.

Here we address an anticipated criticism head-on. This framework trains the model on data curated by skilled researchers; if what we seek to exclude is the preconceptions of skilled practitioners, is that not a contradiction? This study answers with the distinction between "domain knowledge" and "analyst bias." The former is knowledge accumulated across individual projects that is necessary for reading consumers' words (e.g., the difference in meaning between "high price" and "premium feel"). The latter is the prior expectation brought into an individual analytical project—"this time the results should turn out this way." CPT and SFT internalize only the former: the training data contain no project-specific hypotheses, and the SFT data pass mutual review by multiple experts (Section 4.1) to dilute individual idiosyncrasies. Furthermore, whether this distinction actually functions is not left as a matter of belief; it is posed as a measurable claim through the three falsification paths presented in Section 6.5.

### 4.5 The Interdisciplinary Positioning of “Analyst-Bias-Free”

Stated more precisely, this study's "analyst-bias-free" is "hypothesis-generating." Following the Peircean abductive tradition of Section 2.3, it produces consumer insights, their relationships, and strategic implications from observed data patterns, without being constrained by hypotheses fixed before the data are examined. The analyst intervenes at, broadly, two points only: "setting the analytical objective" and "evaluating the output and refining it toward implementation" (Section 6.3.1 decomposes these two into four operational functions). These two are sanctuaries granted to humans alone; an analysis in which they remain undefined is meaningless, however accurate the system (model) may be. It is precisely because of human decision-making that the structuring of data meaning represented in the domain-specialized model's embedding space begins to have value.

The philosophical basis for naming this study "analyst-bias-free" rather than "hypothesis-free" lies in the Peircean logic of abduction as organized by Inoue [32]. The formula of abduction is isomorphic with the KJ method's question—"what label best explains this group of cards?" Abduction is active inference that generates the most probable explanation from observed data, so the name "Hypothesis-Free" would be misleading. Two implications follow: (1) the KJ method is a process in which humans perform the logical form of abduction, and an LLM with that capability can be its executing agent; and (2) "analyst-bias-free" is not a matter of degree but a design that structurally excludes from the analytical process the "established theories," "industry standards," and "validation criteria" of the industry to which the analyst belongs.

To restate what Computational KJ-Hō is: it is a fusion of the knowledge that giants in different practical and scholarly domains have built up—the qualitative-research practice of marketing research, developed largely in industry; the "KJ method," originating with a cultural anthropologist; "abduction," originating with a logician; "Grounded Theory," developed in qualitative research; and large language models, still evolving in artificial intelligence and information processing. It is a challenge to "interdisciplinarity" [24] that crosses disciplines and journal communities, and a study that, while carrying out analysis unbound by any particular "validation criteria," aims to establish standards for judging the "validity" of the knowledge it produces.

## 5 The Evaluation Structure

### 5.1 Why Existing Metrics Are Insufficient

Evaluating qualitative analysis poses challenges that standard NLP metrics cannot address. The Bilingual Evaluation Understudy (BLEU) and the Recall-Oriented Understudy for Gisting Evaluation (ROUGE) measure word-level similarity or recall between generated and reference text; they fit translation or summarization, but each has limits.

First, similarity may fail to fairly evaluate a sentence that is grammatical but meaningless, or vice versa. It tends to score synonyms and other surface variations low, and it scarcely accounts for context and expressive quality arising from word order. Recall likewise tends to underscore reproductions phrased differently, and a lack of semantic and contextual understanding has been noted.

These metrics cannot judge whether the consumer insights this study aims to generate are valuable as a starting point for downstream processes (strategy construction and tactical planning). For example, a high ROUGE score against a reference insight may miss the strategic implication behind it; conversely, a truly novel insight may score low on ROUGE because it deviates from the reference.

Pi et al. [51] presented a five-dimensional evaluation framework in LOGOS, assessing codebook quality along "Reusability," "Descriptive Fitness," "Descriptive Coverage," "Parsimoniousness," and "Consistency"—a major advance over surface-level NLP metrics, well suited to evaluating the analytical structure and results of a codebook. In this study it corresponds to evaluating the "intermediate" analytical structure, asking whether categories are well-formed, consistent, and comprehensive. It cannot, however, address three properties essential to evaluating the quality of marketing research and consumer insights, namely:

(1) whether a consumer insight is actionable in business decision-making;
(2) whether the system, drawing on domain-specific knowledge, distinguishes consumer "attitudes" that tend to be judged conceptually close but should be considered different in deep psychology; and
(3) whether the system discovers consumer insights that human analysts missed.

That is, the evaluation of the quality and usefulness of the final consumer-insight output. This study proposes two novel metrics: (1) InsightExtraction-F1 and (2) MarketingQA. They stand in a complementary relation to LOGOS's evaluation: LOGOS's metrics evaluate the analytical structure of intermediate representations, while the present metrics evaluate final output quality and usefulness. Grasping the quality of the whole system requires both dimensions—akin to distinguishing index quality from search-result quality in information retrieval. We recommend that research on automated qualitative-analysis systems adopt evaluation at both levels.

### 5.2 InsightExtraction-F1

InsightExtraction-F1 is a precision-recall metric for how well the system's consumer insights correspond to a "gold standard." The gold standard is a set of consensus reference insights that skilled marketing researchers produce by individually analyzing the same interview corpus with conventional qualitative methods and then integrating through structured discussion. Because no directly comparable prior work exists, this is an original metric defined and measured for the first time in this study.

Precision is the proportion of system-generated insights that match at least one gold-standard insight above a threshold; recall is the proportion of gold-standard insights that match at least one system-generated insight; F1 is their harmonic mean.

Note that, because this metric measures agreement with a gold standard, the criticism directed at ROUGE in Section 5.1—that a truly novel insight scores low because it deviates from the reference—can apply to it as well. We therefore do not mechanically count as errors those system-generated insights that match no gold-standard item. Skilled practitioners adjudicate whether each is "a valid discovery that humans had missed" or "an unsupported output"; the former are excluded from the precision denominator and passed to the novelty evaluation of Section 5.4. In short, InsightExtraction-F1 measures whether the system reaches the human level, while the human evaluation of Section 5.4 measures whether it makes discoveries beyond the human level—an explicit division of labor.

Before computing F1, the set of system-generated insights is checked against the gold standard by converting each sentence into a semantic vector and using cosine similarity (0–1) between vectors. Using the domain-specialized embeddings that generate the insights to also compute similarity would create "self-grading" (circular evaluation) and inflate results. To prevent this, similarity is computed using an independent frozen general-purpose embedding model (e.g., multilingual-e5-large) as a neutral space, consistently across all conditions. The matching threshold is an independent parameter whose role differs from that of the F1 target. Neither F1 nor the threshold has an externally referenceable established value—both depend on this study's evaluation setting. Rather than fixing the threshold arbitrarily, skilled practitioners assign "match/non-match" labels to a set of insight pairs, and the threshold maximizing agreement between human judgment and mechanical similarity is adopted. The starting point is 0.70, but this is only an initial calibration value; a sensitivity analysis over 0.6–0.8 confirms that F1 is stable.

For F1, we set 0.65 as a minimum feasibility level and 0.70 as the primary research target. These are not direct criteria from prior work, but normative targets set against corroborating evidence that domain specialization is effective. Lee et al. [43] showed on ChemProt [41] that, against an SOTA F1 of 0.64 and a general BERT F1 of 0.74, domain-specialized BioBERT reached up to F1=0.76. Prescott et al. [52] showed that general-purpose generative AI agreed with humans on 71% (5 of 7 themes) in inductive thematic analysis. As both differ from this study in task, they are not direct benchmarks but contextual grounds supporting the effectiveness and attainability of domain specialization.

In threshold calibration, it is a precondition that the frozen embedding model used as the comparison space can represent the semantic differences of this domain's consumer insights at sufficient resolution. If human judgments and mechanical similarity systematically diverge, this should be treated as a problem of the embedding model's suitability—not the threshold—and the comparison space must be reselected.

### 5.3 The MarketingQA Benchmark

MarketingQA is designed as a domain-specific Japanese-language benchmark of an LLM's marketing-research knowledge, following precedents in other fields; as no directly comparable prior work exists, it is an original metric defined and measured for the first time in this study. Kasai et al. [35] developed IgakuQA, a multiple-choice benchmark using Japan's National Medical Licensing Examination, and Hirano [30] developed japanese-lm-fin-harness, a similar benchmark for the Japanese financial domain. MarketingQA applies the same design logic: a set of multiple-choice questions whose structure encompasses the following four context domains—marketing-research methodology, consumer-behavior theory, qualitative-analysis methodology, and strategic marketing reasoning—measuring how far a model has internalized domain knowledge through CPT. The target is an accuracy of 0.75 or higher for the CPT-trained model (like the targets in Section 5.2, a normative goal rather than an empirically established criterion), with the pre-CPT base model's performance as the baseline for the improvement.

MarketingQA is itself a publishable research output and should be released as an open benchmark to demonstrate the reproducibility of a marketing-research domain-specialized LLM. As a guideline for obtaining baselines, pre-CPT general-purpose LLMs are placed as comparators and their performance on MarketingQA is measured. This is similar to the baseline designs of Pi et al. [51], Wen et al. [65], and Wang, Q. et al. [62], and aims to measure the improvement from domain specialization reproducibly.

Further, at the core of this study's evaluation design is an ablation comparing three conditions—"general-purpose LLM," "CPT only," and "CPT+SFT." That is, MarketingQA and InsightExtraction-F1 are measured for (1) a general-purpose LLM, (2) a CPT-trained model, and (3) a CPT+SFT model. This separates and quantifies the effect of CPT (domain-knowledge internalization) and SFT (acquisition of task-specific analytical operations). Prior domain-specialized LLM research often evaluates CPT and SFT together, leaving little evidence that separates the two; this study is also distinctive in giving an empirical answer to this gap.

The concrete baseline values and the performance of the post-CPT model will be reported in this study's "empirical-phase paper."

### 5.4 The Human-Expert Evaluation Protocol and Metrics

Quantitative metrics are necessary but not sufficient for the quality of the consumer insights Computational KJ-Hō produces. A third evaluation component—evaluation by human experts of properties that current automatic metrics cannot capture—is therefore needed. The evaluation protocol adopts a blind-comparison design: skilled marketing researchers with no prior exposure to the qualitative data fed to the system, and blind to whether each consumer insight was system-generated or produced by human analysis, rate the following.

① Usefulness (practically beneficial; five-point scale)
② Actionability (adoptable in practice; five-point scale)
③ Novelty (a new discovery not contained in the human analysts' output)

The "novelty" metric must be handled with particular care, as prior findings are not uniform. Wang, Q. et al. [62], in a pilot study (14 researchers) of inductive thematic analysis using GPT, showed that 14.81% (8 of 54) of the initial codes GPT generated had not been discovered by the human analysts—because GPT read all the data cross-sectionally and extracted patterns. On the other hand, Deiner et al. [17] report that the LLM did not reach the quality or consistency of a human team's qualitative coding grounded in deep insight. Mehta et al. [45], evaluating TA by GenAI (GPT-4o), found it good at identifying comprehensive themes but unable to scrutinize the data deeply or grasp cultural context, and therefore recommend its use in qualitative analysis only as an auxiliary tool. Doshi and Hauser [18] demonstrated a "social dilemma": generative AI raises individual creativity (i.e., novelty) (+8.1%, $p<0.001$), yet the mutual similarity of outputs rises so that collective novelty declines.

This study positions Wang, Q. et al. [62] as strong corroboration while—mindful of its preliminary nature (pilot scale, a specific domain, model generation)—leaving "to what extent a domain-specialized AI can discover insights undetected by humans" as a question for the empirical phase. For "novelty," we adopt Chen et al.'s [14] "Novelty": the proportion of codes discovered only by the LLM. Chen et al. themselves recommend against judging by a fixed threshold, advising that the metric be interpreted in combination with other metrics as a diagnostic tool that prompts qualitative inquiry. This study sets no concrete target for the novelty rate but keeps "novelty" among the key metrics. In light of long practical experience, we do not regard high novelty as unconditionally desirable: a "highly novel consumer insight" may look at first like a "Blue Ocean" (a new market no one has found), when in fact it is merely a market consumers ignore (one with no business potential) that the AI has forcibly extracted—an eventuality that cannot be ruled out. Then again, it may indeed be the early discovery of a Blue Ocean. This is precisely the domain where humans must judge, in the third layer, "strategy generation."

## 6 Discussion

### 6.1 Epistemological Implications: Is the LLM's Processing "Functional Abduction"?

In Section 2.3 we introduced the concept of functional abduction to characterize the output of a domain-specialized LLM that generates consumer insights not predetermined by analytical hypotheses. The concept takes a deliberately modest epistemological stance: it claims a structural equivalence between the LLM's output and abductive reasoning—both novel, data-grounded, and explanatory—but does not claim that the underlying computational process constitutes inference in Peirce's philosophically strict sense.

This study's position is that this question cannot currently be resolved with confidence, and that the practical significance for marketing research is not substantially affected by its resolution. What matters for consumer-insight generation is that output be data-grounded, explanatory, and novel. We note, however, that novelty does not make everything good: human deep psychology is complex, yet it has surprisingly simple regions, and not all of it remains unexplained. Because of this uncertainty, the study requires sustained human oversight. A system that generates views on consumer psychology without human verification would make claims its output cannot independently guarantee.

### 6.2 De-Westernizing AI Epistemology (the WEIRD Problem)

A serious but easily overlooked problem in ongoing scientific research, especially in human psychology and behavioral science, is reliance on WEIRD (Western, Educated, Industrialized, Rich, Democratic) cognition. The validity of taking results from Western, educated, industrialized, rich, democratic samples as "universal human behavioral psychology" is in question [29]. This is not only an academic problem: in practical domains such as "marketing," Western-centric methodologies and criteria are often prioritized. By placing the KJ method—a non-Western integrative technique—at its core, this study demonstrates the de-WEIRDing of Computational KJ-Hō and offers a needed counter-discourse to these established norms.

The computational implementation of the KJ method through CPT specialized to the Japanese-language domain marks an important step toward "de-Westernizing AI epistemology." These two distinctive features are also indispensable for capturing cultural nuance in non-WEIRD markets. Computational KJ-Hō seeks to ensure that the "meaning" AI generates is grounded not in reflections of Western language patterns but in the cultural, linguistic, and professional context tied to the data's origin. Naturally, whether a framework built in Japan—a non-Western environment—works in other countries must be validated; this study will conduct such validation in a non-Western culture and report it in an empirical paper.

### 6.3 Human–AI Collaboration: "Analyst-Bias-Free" Does Not Mean "No Humans Needed"

#### *6.3.1 Redistributing Human–AI Collaboration: The "Human-on-the-Loop" Paradigm*

"Analyst-bias-free" removes a particular form of analyst influence from the analytical loop—the intentional premises that decide which patterns become visible before the data are examined—not the human analyst. It redefines where human judgment applies. In Computational KJ-Hō, the human retains responsibility for the following four functions. These decompose into operational functions the two broad points of intervention stated in Section 4.5 (setting the analytical objective = ① and ②; evaluating the output and refining it toward implementation = ③ and ④).

① Defining the marketing-research objective

② Specifying operational constraints such as brand guidelines and regulatory requirements

③ Evaluating and curating system output

④ Strategic decision-making based on insights
Converting consumer understanding into business action requires human judgment about organizational context, the competitive environment, and strategic prioritization that the system lacks.

This division of labor differs from the model embodied in CollabCoder [25], where the human analyst stays involved throughout the coding process. Computational KJ-Hō removes the human from the coding loop while retaining human involvement in the judgment loop. The analyst's cognitive resources are not absorbed by the labor of processing individual data units but concentrated on the final decisions that require human expertise.

In practice, this redistribution follows a three-step cycle.
First: the human analyst defines the marketing-research objective.
Second: Computational KJ-Hō runs autonomously (from embedding through semantic clustering, insight verbalization, and relational structuring), generating a consumer-insight map without the analyst's intervention.
Third: the analyst evaluates the output (accepting, rejecting, refining, or requesting re-analysis).
This cycle corresponds to the Human-Autonomy Systems Oversight (HASO) model [20], in which the human exercises supervisory control: humans are excluded where analyst bias is most influential—pattern recognition and categorization—and retained in question-setting, interpretation, and decision-making.

#### *6.3.2 Reliability Requirements*
Realizing human-on-the-loop requires system safety and reliability and, above all, "process transparency." If the analyst cannot confirm the reasoning process behind a consumer insight, supervisory control becomes an evidence-free "blind approval." Computational KJ-Hō addresses this through an evidence-link architecture: every element of an output insight presents an explicit connection to the original qualitative data (consumer utterances) from which it was derived. This is an aspect for which the KJ method itself is still valued today, echoing what Kawakita [39] himself advocated as the proper form of scientific method:

① presenting data with the means and methods of acquisition made fully transparent; and
② making the data-processing methods equally transparent.

Being able to check generated consumer insights against the original data provides evidence for auditing and enables data-grounded strategic judgment—an Evidence-Based Marketing Strategy. Downstream stakeholders in the business process cannot reasonably trust content explained only by "because the AI produced it." There is no need to regress here to AI's "black box" problem.

#### *6.3.3 Implications for CSCW Research*
Let us first clarify the relationship between this framework and Computer-Supported Cooperative Work and Social Computing (CSCW). The framework is not a system closed to the interaction between a single analyst and an AI. First, the KJ method at its core is, from its origin, a team-based collaborative methodology in which multiple analysts examine the validity of groupings and diagrams and reach consensus and collective decisions (Section 2.1). Second, the construction and evaluation of the framework are themselves collaborative practices: the SFT training data are curated through mutual review by multiple experts (Section 4.1), and the evaluation gold standard is integrated by multiple skilled researchers through structured discussion (Section 5.2). Third, the output of the third layer functions as a shared basis (a boundary object) on which multiple stakeholders within the organization—researchers, brand managers, and executives—make strategic decisions (Section 6.3.2). What this framework attempts, in short, is not the mere automation of individual work but the redesign of human collaboration within marketing research, an inherently collaborative knowledge practice—squarely within CSCW's central concern: the design of technology to support collaborative work.

On this basis, this study poses three questions worth ongoing discussion to the CSCW community.

First, the problem of automation bias. The risk that humans uncritically accept AI-generated output (Commission Error) has been theoretically formalized in research on automation bias [49]. Prior work has suggested this risk is real and that interface design may be needed to maintain appropriate trust. As the accuracy of generative AI improves rapidly, the human side's critical scrutiny may be reaching its limits. We must accept the twofold problem—that AI has bias and that those who use AI, expert or not, also incur bias—and constantly devise countermeasures.

Second, the problem of the optimal allocation of human–AI collaboration. This study's allocation assigns all pattern recognition to the model and all evaluation and decision-making to the human, but this binary split may not be optimal in every context. Collaborative approaches in which the analyst intervenes at an intermediate stage—e.g., merging or splitting clusters before insight verbalization—may improve analytical quality and merit empirical investigation.

Third, the problem of calibrating trust across human expertise levels. Skilled marketing researchers and junior researchers differ in the skill level required for supervisory control. Moreover, they are experts in marketing research, not in computing; domain experts and computing experts hone their expertise in separate fields, with little likelihood of overlap. If human–AI collaboration is to advance across all knowledge work, examining how to coordinate the independence and cooperation of these two kinds of expertise would broaden and deepen CSCW's understanding.

### 6.4 A Rapidly Maturing Research Field: Timing and Strategic Implications

LLM-based qualitative-analysis research is advancing at an exceptional pace; as recorded in Section 3, at least eight major systems appeared between mid-2025 and early 2026. The individual components of the architecture Computational KJ-Hō proposes have each been explored independently—NGT's embeddings and clustering, SFT-TA's fine-tuning, the multi-agent systems of Thematic-LM and Auto-TA. What differentiates this study is their integration: CPT-based domain specialization and a pipeline connecting analysis to actionable strategy. The window to establish priority is narrowing, making this concept paper a strategic necessity.

Yet this rapid pace also confirms the importance of the research direction. Our decision to enter the CSCW community and accelerate publication reflects a broad recognition that the scalability crisis in qualitative research is a problem worth solving at scale. The contribution of bringing a researcher's perspective drawn from long practice, a non-Western methodology, and domain-specialized CPT to this discourse is, we believe, both timely and indispensable.

### 6.5 Limitations and Future Work

This study establishes the foundational architecture for realizing Computational KJ-Hō, yet several limitations remain.

First, the system currently handles mainly text data, but the marketing domain involves much information arising from non-text elements such as semiotic advertising. In the future, integrating a multimodal CPT model will be indispensable for more holistic integration. Furthermore, sharpening sensitivity to "weak signals" in consumer behavior—which may foreshadow future market expansion or major shifts in preferences—calls for further refinement through advanced outlier-detection algorithms and the like.

Beyond these technical extensions, this study marks the beginning of a long-term research trajectory. The objective measurement of "consumer-insight quality" poses serious challenges. The next step is to develop domain-specialized evaluation protocols and new quantitative metrics that assess the "strategic usefulness" and "interpretive depth" of AI-generated insights. Establishing such rigorous criteria aims to develop Computational KJ-Hō into a globally applicable methodology for qualitative inquiry.

Second, this study is designed for the marketing-research domain, so the qualitative data analyzed are mainly consumer utterances (interview data). The domain specialization achieved through CPT on consumer-psychology and -behavior interview corpora is not, I believe, automatically transferable to other qualitative domains such as medicine, education, public policy, or organizational research; each requires its own CPT investment.

Third, the CPT+SFT approach requires substantial domain data: a large corpus for pre-training and expert-curated insight pairs for fine-tuning. In domains with limited available data, the approach may not be reproducible without first investing in a data-curation foundation.

Fourth, the two metrics proposed in Section 5—InsightExtraction-F1 and MarketingQA—are themselves novel and require validation. Their thresholds and accuracy targets are theoretically derived starting points, not empirically established criteria; ongoing empirical research must establish the relationship between these metric values and practical research usefulness.

Fifth, the response to the WEIRD problem. Under the banner of "de-Westernizing AI epistemology," this study placed the KJ method—long used in Japan—at the core engine of the domain-specialized LLM. This is only a first step in maintaining the cultural, linguistic, and professional context tied to the data's origin. The world cannot be described by a binary of "Western" and "non-Western," nor is "non-Western" homogeneous with "Japan." A truly pluralistic methodology exists in as many forms as there are cultures, and while covering them all is unrealistic, that is no justification for taking the "Western" as the default. We should heed the following warning, which points to the invisibility of research published in languages other than English and to AI's reproduction of the hierarchies of knowledge:
"Research published in languages other than English is often regarded as 'invisible'. Artificial intelligence models are largely trained on Western sources; as a result, the knowledge map of tomorrow redraws the hierarchies of yesterday." [2]

Sixth, this study's central claim—"analyst-bias-free"—is itself a hypothesis requiring empirical falsification. The author makes it falsifiable via three routes: (1) comparing the post-CPT embedding-space structure with the pre-CPT one to show that the resolution of domain-specific distinctions improves quantitatively; (2) comparing, by blind evaluation, the outputs of multiple independent analyst groups (a human team, a general-purpose LLM team, and this framework) on the same data to confirm statistically that the framework is not biased toward particular analyst premises; and (3) measuring the output agreement of multiple CPT models independently trained on the same corpus to confirm reproducibility. These are future work; should they fail to show the framework's superiority, the claim will be revised or withdrawn.

## 7 Conclusion

Qualitative research has long operated under a constraint its practitioners accepted as essential: the depth of understanding it produces is inversely proportional to the volume of data it can process. Even a skilled analyst can hold only about 200 data points in productive analytical tension. Beyond that threshold, the intellectual honesty of qualitative inquiry—sensitivity to context, attention to contradiction, openness to the unexpected—yields to cognitive overload and the selective pressure of an attention that cannot attend to everything at once.

This study proposes Computational KJ-Hō as a theoretical framework for overcoming this constraint without abandoning the epistemological commitment that gives qualitative inquiry its value. Its central claim is that the KJ method's fundamental orientation—letting the structure of meaning emerge from the data rather than imposing it from outside—can be computationally realized at scale through a domain-specialized LLM. What is preserved is not a mere procedure but the epistemological attitude itself: the "analyst-bias-free" commitment to face the "speech" of the data consumers produce, not through a filter of prior expectations and assumptions.

This study's framework makes five contributions to the journal communities of Computational Qualitative Analysis and HCI. First, it proposes the theoretical integration of three intellectual traditions—the KJ method, Grounded Theory, and Peirce's abductive reasoning—which converge on a single epistemological commitment of data-driven explanation generation, clarified here as the conceptual foundation of a computational system. Second, it presents a three-layer architecture extending this commitment to data volumes and analytical contexts beyond human cognitive capacity, with CPT-based domain specialization as its technical foundation. Third, it proposes novel metrics to assess insight quality that standard NLP metrics and existing codebook-quality frameworks cannot address. Fourth, it shows that the WEIRD problem extends to qualitative analysis as well and places a non-Western qualitative methodology at its intellectual core. Fifth, from nearly three decades of global practical experience it derives five practice-grounded problem formulations that existing systems cannot resolve, and translates them into design requirements that integrate domain expertise with computational methodology.

The long-term significance of this study is that the proposed framework may, in the future, extend beyond marketing research. Organizations of every size and field face the same structural challenge: available qualitative data contain understanding that quantitative methods cannot fully capture, yet the resources needed for analysis have historically been limited to organizations with the capacity to invest. A validated Computational KJ-Hō would make "analyst-bias-free" large-scale insight extraction accessible to far broader research contexts. What is democratized is not the already-abundant data, but the analytical infrastructure needed to convert data into understanding.

To restate the central claim: the wisdom of "letting the data speak," which the KJ method has cultivated over 60 years, can now be computationally realized at scale through a domain-specialized LLM. This is not mere efficiency gain but a methodological extension that preserves the epistemological commitment of qualitative research while overcoming its scalability constraint.

The five contributions of this paper ultimately converge on a single possibility: democratizing consumer-insight extraction through computational qualitative methods, so that even organizations without large research teams can access professional-quality consumer understanding. With this prospect, we conclude.